\documentclass[11pt]{article}

\usepackage{epsf}

\newcommand{\Ta}{H_T}
\newcommand{\Tb}{\tilde{H}_T}
\newcommand{\Tc}{E_T}
\newcommand{\Td}{\tilde{E}_T}

\begin{document}


\begin{flushright}
DESY--01--009 \\
hep-ph/0101335 \\
January 2001 \\
\end{flushright}

\vspace{0.8cm}

\begin{center}
{\Large GENERALIZED PARTON DISTRIBUTIONS \\[0.5\baselineskip]
        WITH HELICITY FLIP}

\vspace{2cm}

M. Diehl \\
\textit{Deutsches Elektronen-Synchroton DESY, 22603 Hamburg, Germany}
\end{center}

\vspace{2\baselineskip}

\begin{center}
\textbf{Abstract}
\end{center}

\parbox{0.9\textwidth}{We show that for both quarks and gluons there
are eight generalized parton distributions in the proton: four which
conserve parton helicity and four which do not. We explain why time
reversal invariance does not reduce this number from eight to six, as
previously assumed in the literature.}

\vspace{3\baselineskip}


\section{Introduction}
\label{sec:intro}

The concept of generalized parton distributions
\cite{Muller:1994fv,Ji:1997ek} has recently generated considerable
interest. On the theory side it has become clear that these quantities
contain a wealth of information about the quark and gluon structure of
the nucleon, well beyond what can be learned from the usual parton
densities. From an experimental point of view, the measurement of the
exclusive processes where generalized distributions occur is becoming
possible. There are for instance encouraging preliminary analyses of
deeply virtual Compton scattering, $e p \to e \gamma p$, at DESY
\cite{Saull:1999kt}.

In this note we will be concerned with generalized parton
distributions that describe quark or gluon helicity flip. In the case
of quarks they generalize the usual quark transversity distributions
\cite{Ralston:1979ys,Jaffe:zw}, which remain the least well-known of
the spin densities in the nucleon and are the object of intense
studies. Unfortunately, no process is known to date where the
generalized quark transversity distributions contribute---initial
hopes to access them in vector meson electroproduction
\cite{Collins:1997fb} have not been borne out since the corresponding
hard scattering kernels vanish for symmetry reasons
\cite{Mankiewicz:1998uy}.

The situation is different for gluons: generalized distributions
describing gluon helicity flip appear in deeply virtual Compton
scattering to order $\alpha_s$, and their contribution to the cross
section can be isolated from suitable angular correlations in the
final state \cite{Diehl:1997bu,Hoodbhoy:1998vm,Belitsky:2000jk}. An
important feature of gluon helicity flip distributions is that they do
not mix with quark distributions under evolution. In this sense they
probe gluons in a qualitatively different way than the usual gluon
distributions do.

In ordinary parton distributions, gluon helicity flip can only occur
for targets with spin 1 or higher \cite{Jaffe:1989xy}, since the
change of helicity on the parton side must be compensated by a
corresponding change for the target in order to ensure angular
momentum conservation. There is no such constraint for generalized
distributions, because they admit a transfer of transverse momentum
and thus of orbital angular momentum. The generalized gluon helicity
flip distributions for a spin $\frac{1}{2}$ target thus involve the
orbital angular momentum between partons in an essential way.

The generalized quark and gluon helicity flip distributions for a
hadron with spin $\frac{1}{2}$ have been classified and investigated
in a paper by Hoodbhoy and Ji \cite{Hoodbhoy:1998vm}. By an argument
based on the counting of helicity amplitudes they concluded that in
addition to the four well-known quark helicity conserving
distributions there are \emph{two} quark helicity changing ones, and a
corresponding number for gluon distributions. In this note we wish to
point out that there are actually \emph{four} helicity flip
distributions for quarks and \emph{four} for gluons, i.e., twice as
many as introduced in \cite{Hoodbhoy:1998vm}. In
Section~\ref{sec:quark} we will introduce a complete set of helicity
flip distributions for quarks and discuss their general symmetry
properties. In the following section, we will give a helicity
representation of these distributions and show at which point the
counting argument of Hoodbhoy and Ji fails. We introduce the helicity
flip distributions for gluons in Section~\ref{sec:gluon}, and briefly
discuss their phenomenology in the Compton process in
Section~\ref{sec:compton}. We summarize our findings in
Section~\ref{sec:sum}. In an Appendix we give some technical details
on the form factor decomposition underlying our definition of the
helicity flip distributions. Throughout our paper we will only
consider parton distributions of twist 2.

\section{Quark helicity flip distributions}
\label{sec:quark}

Although the principal practical interest in generalized helicity flip
distributions is at present in the gluon sector, we will first discuss
the case of quarks, where the algebra is slightly less involved. To
introduce our notations, let us first recall the definitions of the
quark helicity conserving distributions~\cite{Hoodbhoy:1998vm},
\begin{eqnarray}
  \label{no-flip-quark}
\lefteqn{ \frac{1}{2} \int \frac{d z^-}{2\pi}\, e^{ix P^+ z^-}
  \langle p',\lambda'|\, 
     \bar{\psi}(-{\textstyle\frac{1}{2}}z)\, 
     \gamma^+ \psi({\textstyle\frac{1}{2}}z)\, 
  \,|p,\lambda \rangle \Big|_{z^+=0,\, \mathbf{z}_T=0} } 
\nonumber \\
&=& \frac{1}{2P^+} \bar{u}(p',\lambda') \left[
  H^q\,  \gamma^+  +
  E^q\, \frac{i \sigma^{+\alpha} \Delta_\alpha}{2m}
  \right] u(p,\lambda) ,
\nonumber \\
\lefteqn{ \frac{1}{2} \int \frac{d z^-}{2\pi}\, e^{ix P^+ z^-}
  \langle p',\lambda'|\, 
     \bar{\psi}(-{\textstyle\frac{1}{2}}z)\, 
     \gamma^+ \gamma_5\, \psi({\textstyle\frac{1}{2}}z)\, 
  \,|p,\lambda \rangle \Big|_{z^+=0,\, \mathbf{z}_T=0} } 
\nonumber \\
&=& \frac{1}{2P^+} \bar{u}(p',\lambda') \left[
  \tilde{H}^q\, \gamma^+ \gamma_5 +
  \tilde{E}^q\, \frac{\gamma_5 \Delta^+}{2m}
  \right] u(p,\lambda) ,
\end{eqnarray}
where $p, p'$ and $\lambda, \lambda'$ respectively denote proton
momenta and helicities. We use light-cone coordinates $v^\pm = (v^0
\pm v^3)/\sqrt{2}$ and $\mathbf{v}_T = (v^1, v^2)$ for any four-vector
$v$, Ji's kinematical variables
\begin{equation}
  \label{kin-def}
P = \frac{1}{2}\, (p+p') , \qquad
\Delta = p'-p , \qquad
\xi= \frac{p^+-p'^+}{p^+ +p'^+} ,
\end{equation}
and $t = \Delta^2$. Throughout this paper we will work in the
light-cone gauge $A^+=0$, so that no gauge link appears between the
quark field operators in Eq.~(\ref{no-flip-quark}).

The quark helicity flip distributions go with the Dirac matrix
$\sigma^{+i}$, where $i=1,2$ is a transverse index, and we define
\begin{eqnarray}
  \label{flip-quark}
\lefteqn{ \frac{1}{2} \int \frac{d z^-}{2\pi}\, e^{ix P^+ z^-}
  \langle p',\lambda'|\, 
     \bar{\psi}(-{\textstyle\frac{1}{2}}z)\, i \sigma^{+i}\, 
     \psi({\textstyle\frac{1}{2}}z)\, 
  \,|p,\lambda \rangle \Big|_{z^+=0,\, \mathbf{z}_T=0} } 
\nonumber \\
&=& \frac{1}{2P^+} \bar{u}(p',\lambda') \left[
  \Ta^q\, i \sigma^{+i} +
  \Tb^q\, \frac{P^+ \Delta^i - \Delta^+ P^i}{m^2} \right.
\nonumber \\
&& \left. \hspace{5em} {}+
  \Tc^q\, \frac{\gamma^+ \Delta^i - \Delta^+ \gamma^i}{2m} +
  \Td^q\, \frac{\gamma^+ P^i - P^+ \gamma^i}{m}
  \right] u(p,\lambda).
\hspace{2em}
\end{eqnarray}
$\Ta^q$ and $-\Tc^q$ respectively correspond to the distributions
$H_{Tq}$ and $E_{Tq}$ introduced by Hoodbhoy and Ji
\cite{Hoodbhoy:1998vm}, whereas the distributions $\Tb^q$ and $\Td^q$
are new. Eq.~(\ref{flip-amplitudes}) below explicitly shows that the
four Dirac structures on the right-hand side of Eq.~(\ref{flip-quark})
are linearly independent. On the other hand, there cannot be more than
four distributions to parameterize the left-hand side of
Eq.~(\ref{flip-quark}). We have two quark-antiquark operators (one
with $i=1$ and one with $i=2$) and four helicity combinations
($\lambda, \lambda'$) of the proton states. With the constraints of
parity invariance these $2 \times 4 = 8$ matrix elements are related
pairwise, so that the number of independent distributions is four. The
functions $\Ta^q$, $\Tb^q$, $\Tc^q$, and $\Td^q$ thus represent a
complete set of generalized quark helicity flip distributions.

In the definition of quark transversity one often uses the matrix
$\sigma^{+j}\gamma_5$ with $j=1,2$ instead of $\sigma^{+i}$. With the
relation
\begin{equation}
  \label{sigma-five}
\sigma^{\alpha\beta}\gamma_5 = - \frac{i}{2}
\epsilon^{\alpha\beta\gamma\delta}\, \sigma_{\gamma\delta} ,
\end{equation}
where our convention is $\epsilon_{0123} = 1$, we find that the
definition (\ref{flip-quark}) is equivalent to
\begin{eqnarray}
  \label{flip-quark-alt}
\lefteqn{ \frac{1}{2} \int \frac{d z^-}{2\pi}\, e^{ix P^+ z^-}
  \langle p',\lambda'|\, 
     \bar{\psi}(-{\textstyle\frac{1}{2}}z)\, \sigma^{+j}\gamma_5\, 
     \psi({\textstyle\frac{1}{2}}z)\, 
  \,|p,\lambda \rangle \Big|_{z^+=0,\, \mathbf{z}_T=0} } 
\nonumber \\
&=& \frac{1}{2P^+} \bar{u}(p',\lambda') \left[
  \Ta^q\, \sigma^{+j}\gamma_5 +
  \Tb^q\, \frac{\epsilon^{+j\alpha\beta}\, \Delta_\alpha P_\beta}{m^2}
  \right.
\nonumber \\
&& \left. \hspace{5em} {}+
  \Tc^q\, \frac{\epsilon^{+j\alpha\beta}\, \Delta_\alpha
                          \gamma_\beta}{2m} +
  \Td^q\, \frac{\epsilon^{+j\alpha\beta}\, P_\alpha \gamma_\beta}{m}
  \right] u(p,\lambda) .
\end{eqnarray}

The counting argument in \cite{Hoodbhoy:1998vm} was based on time
reversal invariance. Let us therefore see what this symmetry implies
for the distributions we have introduced. Introducing the antiunitary
operator $V$ that implements time reversal in Hilbert space, we can
insert $1 = V^{-1}\, V$ in the matrix elements on the left-hand sides
of Eqs.~(\ref{no-flip-quark}), (\ref{flip-quark}) and obtain
\begin{equation}
  \label{time}
f^q(x,\xi,t) = f^q(x,-\xi,t)
\end{equation}
for $f = H$, $\tilde{H}$, $E$, $\tilde{E}$, $\Ta$, $\Tb$, $\Tc$, and
\begin{equation}
  \label{special-time}
\Td^q(x,\xi,t) = - \Td^q(x,-\xi,t) .
\end{equation}
That time reversal changes the sign of $\xi$ reflects the fact that
under time reversal initial and final states are interchanged. Taking
the complex conjugates of Eqs.~(\ref{no-flip-quark}) and
(\ref{flip-quark}) gives, on the other hand,
\begin{equation}
  \label{hermit}
\Big[f^q(x,\xi,t)\Big]^* = f^q(x,-\xi,t)
\end{equation}
for all distributions except $\Td^q$, and
\begin{equation}
  \label{special-hermit}
\Big[\Td^q(x,\xi,t)\Big]^* = - \Td^q(x,-\xi,t) .
\end{equation}
Taking these constraints together we see that all 8 distributions are
required to be real valued as a consequence of time reversal
invariance. In other words, this symmetry fixes the phases of the
distributions, but does not require any linear combination of them to
be zero.

That $\Tc^q$ and $\Td^q$ have opposite behavior under time reversal,
as borne out by Eqs.~(\ref{time}) and (\ref{special-time}), could have
been anticipated from inspection of the tensors that multiply them in
their definition~(\ref{flip-quark}). Namely, $\Delta$ changes sign
under $p \leftrightarrow p'$ but $P$ does not. As we have seen, this
does not constrain either of these distributions to be zero. It is
interesting to note that this situation changes if instead of the
bilocal quark-antiquark operator in Eq.~(\ref{flip-quark}) one
considers the local one,
\begin{equation}
  \label{local-matrix}
T^{\alpha\beta} = 
  \langle p',\lambda'|\, 
     \bar{\psi}(0)\, i \sigma^{\alpha\beta}\, \psi(0)\, 
  \,|p,\lambda \rangle .
\end{equation}
In the Appendix we will show that, under the constraints of parity
invariance, a complete set of Dirac bilinears $\bar{u}(p',\lambda')\,
\Gamma^{\alpha\beta}\, u(p,\lambda)$ to define the form factor
decomposition of the matrix element~(\ref{local-matrix}) is given by
the four bilinears on the right-hand side of
Eq.~(\ref{flip-quark}). Time reversal now does imply that the form
factor multiplying the fourth bilinear, $\bar{u} (\gamma^\alpha
P^\beta - P^\alpha \gamma^\beta) u$, must vanish.

The components $(\alpha,\beta)=(+,i)$ of the tensor
(\ref{local-matrix}) are readily obtained by integrating the left-hand
side of Eq.~(\ref{flip-quark}) over $x$ and multiplying with $2
P^+$. The corresponding first $x$-moments of the quark distributions
are thus just the form factors of the local matrix element, and by
Lorentz invariance only depend on the invariant transfer $t$. That the
form factor corresponding to $\Td^q$ must vanish can thus be directly
seen by integrating the relation (\ref{special-time}) over its support
$-1 \le x \le 1$. Since the result must not depend on $\xi$, one finds
in fact
\begin{equation}
  \label{moment}
\int_{-1}^1 dx\, \Td^q(x,\xi,t) = 0 .
\end{equation}

Thus, we have found that by time reversal invariance there are only
\emph{three} independent form factors of local matrix element
(\ref{local-matrix}) but \emph{four} independent generalized quark
distributions to describe the bilocal matrix element
(\ref{flip-quark}). In other words, the first moment of $\Td^q$ is
zero by time reversal symmetry, but not its higher moments, $\int dx\,
x^{n-1}\, \Td^q(x,\xi,t)$ with $n>1$. In fact, these correspond to
local matrix elements as in (\ref{local-matrix}) but with further
additional derivatives $\partial^+$. The corresponding Lorentz tensors
have rank larger than two and allow more than 3 independent form
factors. For more detail we refer to the Appendix.

\section{Helicity representation}
\label{sec:helicity}

To investigate the spin structure of the generalized parton
distributions it is useful to represent them in a form similar to that
of helicity amplitudes. Since we are dealing here with matrix elements
involving two independent proton momenta, some comments are in order
regarding the choice of helicity states for the protons. We note that
in the definitions of the distributions one singles out a direction
that defines the light-cone coordinates (in a physical process where
these distributions appear, this direction is provided by the hard
probe, such as the virtual photon in deeply virtual Compton
scattering). It is useful to also utilize this light-cone direction
for defining the spin states for the protons with momenta $p$ and
$p'$. This leads to the concept of light-cone helicity states
\cite{Kogut:1970xa}.\footnote{For a brief review of their
construction, cf.\ e.g.~\cite{Diehl:2000xz}. The results of that paper
also illustrate how light-cone helicity naturally appears in the
context of generalized parton distributions.} A set of corresponding
spinors, given in the usual Dirac representation,
is~\cite{Brodsky:1998de}
\begin{eqnarray}
  \label{LC-spinors}
u_{\mathrm{LC}}(p,+) &=& \frac{1}{\sqrt{2(p^0 + p^3)}} 
     \left( \begin{array}{c} 
      p^0 + p^3 + m \\ p^1 + i p^2 \\ p^0 + p^3 - m \\ p^1 + i p^2
     \end{array} \right) ,
\nonumber \\
u_{\mathrm{LC}}(p,-) &=& \frac{1}{\sqrt{2(p^0 + p^3)}} 
     \left( \begin{array}{c} 
      - p^1 + i p^2 \\ p^0 + p^3 + m \\ p^1 - i p^2 \\ - p^0 - p^3 + m
     \end{array} \right) 
\end{eqnarray}
for a particle with mass $m$. For the sake of legibility we denote
helicity labels for fermions by $+$ and $-$ instead of $+\frac{1}{2}$
and $-\frac{1}{2}$ here are in the following. For zero mass the
spinors (\ref{LC-spinors}) are identical with the usual helicity
spinors $u_{\mathrm{H}}(p,\pm)$, but not if the mass is finite. Since
in phenomenological applications one will often deal with usual
helicity amplitudes let us briefly give the transformation between the
two sets. It can be written as
\begin{equation}
\left( \begin{array}{r} u_{\mathrm{H}}(p,+) \\ 
                        u_{\mathrm{H}}(p,-) \end{array} \right) = \\
{\mathbf{U}} \cdot
\left( \begin{array}{r} u_{\mathrm{LC}}(p,+) \\ 
                        u_{\mathrm{LC}}(p,-) 
       \end{array} \right)
\end{equation}
with a unitary matrix
\begin{equation}
{\mathbf{U}} = N^{-1}\,
\renewcommand{\arraystretch}{1.4}
\left( \begin{array}{rr} 
       (|\mathbf{p}| + p^3) \sqrt{p^0 + |\mathbf{p}|}  &
       (p^1 + i p^2) \sqrt{p^0 - |\mathbf{p}|}        \\
     - (p^1 - i p^2) \sqrt{p^0 - |\mathbf{p}|}         &
       (|\mathbf{p}| + p^3) \sqrt{p^0 + |\mathbf{p}|}
       \end{array} \right) , \\
\end{equation}
where
\begin{equation}
N = \sqrt{2(p^0 + p^3)\, |\mathbf{p}|\, (|\mathbf{p}| + p^3)} .
\end{equation}
We see that for a right-moving particle ($p^3 > 0$) the ratio $|U_{+-}
/U_{++}| = |U_{-+} /U_{--}|$ between off-diagonal and diagonal
elements in this matrix is bounded by $m /(2 |\mathbf{p}|)$. This
means that in reference frames where the particle moves fast to the
right, the difference between usual and light-cone helicity is
small. In following we will only use light-cone spinors and drop the
subscript $\mathrm{LC}$, and understand ``helicity'' as ``light-cone
helicity''.

\begin{figure}
   \begin{center}
        \leavevmode
        \epsfxsize=0.99\hsize
        \epsfbox{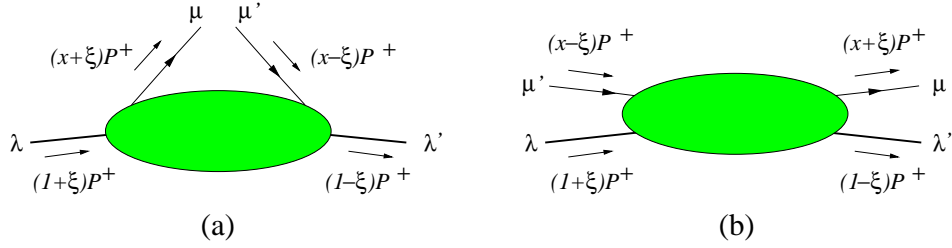}
   \end{center}
\caption{\label{fig:amps} Representation of a generalized parton
distribution in the region $\xi<x<1$. The flow of plus-momentum is
indicated by arrows, and the labels $\lambda$, $\lambda'$, $\mu$,
$\mu'$ denote helicities. (a) shows the ordering of lines as ``proton
in, quark out, quark back in, proton out'' that is common for parton
distributions. (b) displays the order ``proton in, quark in, quark
out, proton out'' appropriate for a scattering amplitude.}
\end{figure}

Let us now discuss the helicity of the partons. It is often said that
parton distributions can be represented as amplitudes for the
scattering of a parton on a proton, see Fig.~\ref{fig:amps}. This
observation is at the base of the covariant parton model
\cite{Landshoff:1971ff}. It will be important in our context that the
above formulation is somewhat imprecise: rather, parton distribution
are amplitudes that are integrated over the minus- and transverse
momentum of the partons.\footnote{Due to the integration over the
parton minus-momentum they are both amplitudes and discontinuities of
amplitudes. In other words, it does not matter whether the field
operators in their definition are time ordered or not
\protect\cite{Jaffe:1983hp}.} This is readily seen by rewriting the
Fourier transform occurring in the definitions of our distributions as
\begin{eqnarray}
  \label{off-shell}
\lefteqn{ \int \frac{d z^-}{2\pi}\, e^{ix P^+ z^-} 
\langle p',\lambda'|\, {\cal O}(z)
  \,|p,\lambda \rangle \Big|_{z^+=0,\,\mathbf{z}_T=0} } 
\nonumber \\
&=& \int \frac{dk^-\, d^2 k_T}{(2\pi)^4}
\left[ \int d^4z\, e^{i k\cdot z}\,
       \langle p',\lambda'|\, {\cal O}(z) \,|p,\lambda \rangle 
     \right]_{k^+ = x P^+} ,
\end{eqnarray}
where ${\cal O}(z)$ stands for the relevant quark-antiquark
operators. The expression in square brackets corresponds to the
discontinuity of an amplitude with off-shell quark legs. This
off-shellness is integrated over, keeping only the parton
plus-momentum fixed.

In order to assign helicities to the quarks it is convenient to have
them on-shell. A way to achieve this is to quantize the theory on the
light-cone and to work in a noncovariant, Hamiltonian
framework~\cite{Kogut:1970xa,Brodsky:1998de}. In this framework, the
dynamically independent (so called ``good'') part of the fermion field
is given by $\phi(z) = P_+ \psi(z)$ with the projector $P_+ =
\frac{1}{2} \gamma^- \gamma^+$. It can be written as
\begin{equation}
  \label{LC-fields}
\phi(z) = \frac{1}{\sqrt{2}} 
     \left( \begin{array}{r} 
      \phi_R(z) \\ \phi_L(z) \\ \phi_R(z) \\ -\phi_L(z)
     \end{array} \right) ,
\end{equation}
where again we use the Dirac representation of four-spinors. At
light-cone time $z^+=0$, the Fourier components of $\phi_R(z)$ are the
annihilation operator for an on-shell quark with helicity
$+\frac{1}{2}$ and the creation operator for an on-shell antiquark
with helicity $-\frac{1}{2}$. Conversely, the Fourier components of
$\phi_L(z)$ are the annihilation operator for an on-shell quark with
helicity $-\frac{1}{2}$ and the creation operator for an on-shell
antiquark with helicity $+\frac{1}{2}$. The operators occurring in the
definitions of the quark distributions can be written as
\begin{eqnarray}
{\cal O}_{+,+} =
\frac{1}{\sqrt{2}}\, \phi^\dagger_R\, \phi^{\phantom{\dagger}}_R
&=& \frac{1}{4}\, 
  \bar{\psi}\, \gamma^+ (1+\gamma_5)\, \psi ,
\nonumber \\
{\cal O}_{-,-} =
\frac{1}{\sqrt{2}}\, \phi^\dagger_L\, \phi^{\phantom{\dagger}}_L
&=& \frac{1}{4}\, \bar{\psi}\, \gamma^+ (1-\gamma_5)\, \psi ,
\nonumber \\
{\cal O}_{-,+} =
\frac{1}{\sqrt{2}}\, \phi^\dagger_L\, \phi^{\phantom{\dagger}}_R
&=& - \frac{i}{4}\, \bar{\psi}\, \sigma^{+1} (1+\gamma_5)\, \psi
= - \frac{i}{4}\, \bar{\psi}\, (\sigma^{+1} - i \sigma^{+2})\, \psi ,
\nonumber \\
{\cal O}_{+,-} =
\frac{1}{\sqrt{2}}\, \phi^\dagger_R\, \phi^{\phantom{\dagger}}_L
&=& \frac{i}{4}\, 
  \bar{\psi}\, \sigma^{+1} (1-\gamma_5)\, \psi \phantom{-}
= \frac{i}{4}\, \bar{\psi}\, (\sigma^{+1} + i \sigma^{+2})\, \psi ,
\hspace{2em}
\end{eqnarray}
where for brevity we have omitted the position arguments $\pm z/2$ of
the field operators. For definiteness we will in the following
restrict ourselves to the region $\xi<x<1$ of plus-momentum fractions,
where the generalized quark distributions describe the emission of a
quark with plus-momentum $(x+\xi)\, P^+$ and its reabsorption with
plus-momentum $(x-\xi)\, P^+$.

We now are in a position to define the matrix elements
\begin{eqnarray}
  \label{on-shell}
A_{\lambda'\mu', \lambda\mu} &=&
\int \frac{d z^-}{2\pi}\, e^{ix P^+ z^-}
  \langle p',\lambda'|\, {\cal O}_{\mu',\mu}(z)
  \,|p,\lambda \rangle \Big|_{z^+=0,\, \mathbf{z}_T=0} 
\nonumber \\
&=& \int \frac{d^2 k_T}{(2\pi)^3}
\left[\, \int dz^-\,d^2 z_T\, e^{i k\cdot z}\,
       \langle p',\lambda'|\, {\cal O}_{\mu',\mu}(z) \,
       |p,\lambda \rangle\,
     \right]_{z^+=0,\, k^+ = x P^+}
\nonumber \\
\end{eqnarray}
for definite parton helicities $\mu$ and $\mu'$. We remark that our
labeling of the helicities corresponds to the ordering of lines shown
in Fig.~\ref{fig:amps}(a) and is different from the usual one for
helicity amplitudes, represented in Fig.~\ref{fig:amps}(b). Notice in
Eq.~(\ref{on-shell}) that in the noncovariant framework there is no
longer an integration over the minus-momentum of the partons as there
was in Eq.~(\ref{off-shell}). One now keeps the condition $z^+=0$ in
the bilocal operator since it is at that point where the field
operators can be replaced in a simple manner by the creation and
annihilation operators for on-shell partons. Put in a different way,
the $k^-$ of the partons is not integrated over because it is fixed by
the on-shell condition. We emphasize, however, that still one does
integrate over the transverse parton momentum.

As we have discussed, the $A_{\lambda'\mu',\lambda\mu}$ are not
exactly helicity amplitudes. What is important in our context is that
they share several symmetry properties which are satisfied by helicity
amplitudes. In particular, one has the relations
\begin{equation}
  \label{parity}
A_{-\lambda'-\mu', -\lambda-\mu} = (-1)^{\lambda'-\mu'-\lambda+\mu}\,
                                   A_{\lambda'\mu', \lambda\mu}
\end{equation}
from parity invariance, provided one works in a reference frame where
the momenta ${\mathbf{p}}$ and ${\mathbf{p}}'$ lie in the $x$-$z$
plane, which we will do from now on. Explicit calculation gives
\begin{eqnarray}
  \label{no-flip-amplitudes}
A_{++,++} &=& \sqrt{1-\xi^2} \left( \frac{H^q+\tilde{H}^q}{2} - 
            \frac{\xi^2}{1-\xi^2}\, \frac{E^q+\tilde{E}^q}{2} \right) , 
\nonumber \\
A_{-+,-+} &=& \sqrt{1-\xi^2} \left( \frac{H^q-\tilde{H}^q}{2} - 
            \frac{\xi^2}{1-\xi^2}\, \frac{E^q-\tilde{E}^q}{2} \right) , 
\nonumber \\
A_{++,-+} &=& - \epsilon\,
              \frac{\sqrt{t_0-t}}{2m}\, \frac{E^q-\xi\tilde{E}^q}{2} , 
\nonumber \\
A_{-+,++} &=& \epsilon\,
              \frac{\sqrt{t_0-t}}{2m}\, \frac{E^q+\xi\tilde{E}^q}{2} ,
\end{eqnarray}
in the quark helicity conserving sector, and
{\samepage
\begin{eqnarray}
  \label{flip-amplitudes}
A_{++,+-} &=&   \epsilon\, \frac{\sqrt{t_0-t}}{2m} \left( \Tb^q
              + (1-\xi)\, \frac{\Tc^q + \Td^q}{2} \right) , 
\nonumber \\
A_{-+,--} &=&   \epsilon\, \frac{\sqrt{t_0-t}}{2m} \left( \Tb^q
              + (1+\xi)\, \frac{\Tc^q - \Td^q}{2} \right) , 
\nonumber \\
A_{++,--} &=& \sqrt{1-\xi^2} \left(\Ta^q + \
              \frac{t_0-t}{4 m^2}\, \Tb^q -
              \frac{\xi^2}{1-\xi^2}\, \Tc^q +
              \frac{\xi}{1-\xi^2}\, \Td^q \right) , 
\nonumber \\
A_{-+,+-} &=& - \sqrt{1-\xi^2}\; \frac{t_0-t}{4 m^2}\, \Tb^q
\end{eqnarray}
for quark helicity flip, with the other helicity combinations given by
parity invariance. Here
}
\begin{equation}
- t_0 = \frac{4 m^2 \xi^2}{1-\xi^2}
\end{equation}
is the minimum value of $-t$ for given $\xi$, and $\epsilon =
\mathrm{sgn}(D^1)$, where $D^1$ is the $x$-component of $D^\alpha =
P^+ \Delta^\alpha - \Delta^+ P^\alpha$. The case $D^1 = 0$ corresponds
to $t=t_0$ so that no ambiguity appears in
Eqs.~(\ref{no-flip-amplitudes}) and (\ref{flip-amplitudes}) at that
point. Apart from restriction that ${\mathbf{p}}$ and ${\mathbf{p}}'$
lie in $x$-$z$ plane, Eqs.~(\ref{no-flip-amplitudes}) and
(\ref{flip-amplitudes}) are valid for any choice of proton momenta.

We note in Eqs.~(\ref{no-flip-amplitudes}) and (\ref{flip-amplitudes})
that the matrix elements where helicity is not conserved vanish like
\begin{equation}
  \label{prefactor}
\left( \frac{\sqrt{t_0-t}}{2m} \right)^{|\lambda'-\mu'-\lambda+\mu|}
\end{equation}
in the collinear limit $t=t_0$, reflecting the fact that the mismatch
of the helicities has to be compensated by one or two units of orbital
angular momentum in order to ensure angular momentum conservation. One
finds that in the collinear limit the distribution $\Tb^q$ decouples,
whereas in the forward limit, $t=0, \xi=0$, the only nonzero
contribution comes from $\Ta^q$, which in that limit reduces to the
conventional quark transversity distribution, $\Ta^q(x,0,0) = \delta
q(x)$.

From Eq.~(\ref{flip-amplitudes}) we explicitly see that the Dirac
bilinears multiplying the distributions $\Ta^q$, $\Tb^q$, $\Tc^q$,
$\Td^q$ in the decomposition~(\ref{flip-quark}) are linearly
independent. If the right-hand side of (\ref{flip-quark}) is to be
identically zero, then all four helicity combinations in
Eq.~(\ref{flip-amplitudes}) must vanish, and this is only the case if
all four distributions are zero.

Let us now see why the counting argument of Hoodbhoy and Ji
\cite{Hoodbhoy:1998vm} does not apply. They invoke that with the
constraints from parity and time reversal invariance there are only
six independent helicity amplitudes for elastic quark-proton
scattering. This is correct, but closer inspection reveals that three
of these amplitudes flip the helicity of the quark and three do
not. This does not correspond to the four non-flip distributions in
Eq.~(\ref{no-flip-quark}) and the two flip distributions $\Ta^q$ and
$\Tc^q$ considered in Ref.~\cite{Hoodbhoy:1998vm}.

At this point we must remember that the matrix elements defining
generalized parton distributions are \emph{not} helicity amplitudes in
the strict sense. Time reversal exchanges initial and final state
momenta. For usual helicity amplitudes, this exchange can be
compensated for by a suitable rotation in the center-of-mass of the
collision, where the three-momenta of the incoming and outgoing
particles have equal length. The matrix elements (\ref{no-flip-quark})
and (\ref{flip-quark}), however, refer to a fixed light-cone axis with
respect to which transverse, plus- and minus-momenta are
defined. Notably, the integration over transverse parton momenta in
Eq.~(\ref{on-shell}) refers to this particular axis. Time reversal
changes $\xi$ into $-\xi$ and thus provides the relations (\ref{time})
and (\ref{special-time}). One might perform a Lorentz transformation
to another frame in order to compensate for this change, but such a
transformation will not leave the light-cone axis invariant. In the
new frame, the generalized parton distributions are then no longer
given as in Eq.~(\ref{on-shell}), i.e., by an integration over the
transverse parton momentum with all plus-momenta fixed.

One may ask whether it is possible to find further time reversal
constraints, for instance by first considering the matrix elements
that correspond to (\ref{on-shell}) but are not integrated over the
parton $\mathbf{k}_T$, trying to obtain constraints of the type one
has for usual helicity amplitudes, and then performing the
integrations over $\mathbf{k}_T$. We will not pursue this here, but
remark that even for fixed $k^+$ and $\mathbf{k}_T$ one still has
singled out a light-cone direction, namely by the condition $z^+=0$ in
the matrix element (\ref{on-shell}).

The argument just given indicates however that the number of
independent structures \emph{is} reduced from eight to six in the case
$\xi = 0$ (note that this does not imply $p=p'$ since one can still
have $t\neq 0$). Clearly, the change $\xi\to -\xi$ of time reversal is
of no consequence then. Indeed, $\Td^q$ vanishes at $\xi=0$ due to the
constraint (\ref{special-time}). At the same point $\tilde{E}^q$,
although nonzero, decouples from all amplitudes because in its
definition (\ref{no-flip-quark}) it is multiplied with $\Delta^+=-2\xi
P^+$. Inspecting Eqs.~(\ref{no-flip-amplitudes}) and
(\ref{flip-amplitudes}) we then find $A_{++,-+} = -A_{-+,++}$ and
$A_{++,+-} = -A_{+-,++}$, where in the second equation we have used
the parity relation (\ref{parity}). These are precisely the
constraints on usual helicity amplitudes arising from time reversal
invariance. We see that one of the two distributions that have thus
been removed flips quark helicity, while the other does not, in
accordance with our remark above.

The main results of our discussion can be generalized to targets with
arbitrary spin. As is well known, the number of ordinary quark or
gluon distributions is equal to the corresponding number of
independent helicity amplitudes allowed by parity and time reversal
invariance (and by helicity conservation, since one is dealing with
forward amplitudes) \cite{Jaffe:zw}.  For generalized parton
distributions, the counting works differently. Their number is
obtained from counting the helicity amplitudes under the constraints
of parity invariance only. Time reversal invariance determines the
behavior of the generalized distributions under exchange of the hadron
momenta ($\xi\to -\xi$) and fixes their phase.

\section{Gluon helicity flip distributions}
\label{sec:gluon}

We now turn to gluon distributions. As is well known, there are again
four distributions conserving gluon helicity,
\begin{eqnarray}
  \label{no-flip-gluon}
\lefteqn{ \frac{1}{P^+} \int \frac{d z^-}{2\pi}\, e^{ix P^+ z^-}
  \langle p',\lambda'|\, 
     F^{+i}(-{\textstyle\frac{1}{2}}z)\, 
     F^{+i}({\textstyle\frac{1}{2}}z)\, 
  \,|p,\lambda \rangle \Big|_{z^+=0,\, \mathbf{z}_T=0} } \hspace{4em}
\nonumber \\
&=& \frac{1}{2P^+} \bar{u}(p',\lambda') \left[
  H^g\, \gamma^+ +
  E^g\, \frac{i \sigma^{+\alpha} \Delta_\alpha}{2m} 
  \right] u(p,\lambda) ,
\nonumber \\
\lefteqn{ - \frac{i}{P^+} \int \frac{d z^-}{2\pi}\, e^{ix P^+ z^-}
  \langle p',\lambda'|\, 
     F^{+i}(-{\textstyle\frac{1}{2}}z)\, 
          \tilde{F}^{+i}({\textstyle\frac{1}{2}}z)\, 
  \,|p,\lambda \rangle \Big|_{z^+=0,\, \mathbf{z}_T=0} } \hspace{4em}
\nonumber \\
&=& \frac{1}{2P^+} \bar{u}(p',\lambda') \left[
  \tilde{H}^g\, \gamma^+ \gamma_5 +
  \tilde{E}^g\, \frac{\gamma_5 \Delta^+}{2m}\, \, 
  \right] u(p,\lambda) ,
\end{eqnarray}
where $\tilde{F}^{\alpha\beta} = \frac{1}{2}
\epsilon^{\alpha\beta\gamma\delta} F_{\gamma\delta}$ is the dual field
strength tensor and a summation over $i=1,2$ is implied. These
definitions differ from those of Hoodbhoy and Ji
\cite{Hoodbhoy:1998vm}, ours are normalized such that in the forward
limit one has
\begin{equation}
H^g(x,0,0) = x g(x) , \hspace{2em} 
\tilde{H}^g(x,0,0) = x \Delta g(x)
\end{equation}
with the usual spin averaged and spin dependent gluon densities $g(x)$
and $\Delta g(x)$. Compared with Ref.~\cite{Hoodbhoy:1998vm}, we have
\begin{equation}
2 x H_g \Big|_{\mathrm{Ref.~[9]}} = H^g \Big|_{\mathrm{here}} ,
\end{equation}
and analogous relations for $E^g$, $\tilde{H}^g$, and
$\tilde{E}^g$. The gluon helicity flip distributions involve the gluon
tensor operator ${\mathbf S} F^{+i}(-{\textstyle\frac{1}{2}}z)\,
F^{+j}({\textstyle\frac{1}{2}}z)$, where ${\mathbf S}$ denotes
symmetrization in $i$ and $j$ and subtraction of the trace. To
parameterize this structure we use the same Dirac bilinears as in the
definition (\ref{flip-quark}) for quarks and introduce
\begin{eqnarray}
  \label{flip-gluon}
\lefteqn{ - \frac{1}{P^+} \int \frac{d z^-}{2\pi}\, e^{ix P^+ z^-}
  \langle p',\lambda'|\, {\mathbf S}
     F^{+i}(-{\textstyle\frac{1}{2}}z)\,
     F^{+j}({\textstyle\frac{1}{2}}z)
  \,|p,\lambda \rangle \Big|_{z^+=0,\,\mathbf{z}_T=0} } \hspace{1.5em}
\nonumber \\
&=& {\mathbf S}\,
\frac{1}{2 P^+}\, \frac{P^+ \Delta^j - \Delta^+ P^j}{2 m P^+}
\nonumber \\
&\times& \bar{u}(p',\lambda') \left[
  \Ta^g\, i \sigma^{+i} +
  \Tb^g\, \frac{P^+ \Delta^i - \Delta^+ P^i}{m^2} \right.
\nonumber \\
&& \left. \hspace{2.8em} {}+
  \Tc^g\, \frac{\gamma^+ \Delta^i - \Delta^+ \gamma^i}{2m} +
  \Td^g\, \frac{\gamma^+ P^i - P^+ \gamma^i}{m}\, 
   \right] u(p,\lambda) .
\hspace{2em} 
\end{eqnarray}
Our distributions $\Ta^g$ and $\Tc^g$ are identical with $H_G^T$ and
$E_G^T$ used in \cite{Belitsky:2000jk}, and related with those of
Hoodbhoy and Ji by
\begin{equation}
- 2x H_{Tg} \Big|_{\mathrm{Ref.~[9]}} = \Ta^g \Big|_{\mathrm{here}} , 
\hspace{2em}
- 2x E_{Tg} \Big|_{\mathrm{Ref.~[9]}} = \Tc^g \Big|_{\mathrm{here}} .
\end{equation}
As in the case of quarks, one can see that there are at most four
independent distributions to parameterize the matrix element
(\ref{flip-gluon}) with the constraints of parity invariance, and we
have shown in Section~\ref{sec:helicity} that the four Dirac bilinears
are linearly independent.

The support of all eight gluon distributions is $-1 \le x \le 1$. It
is easy to see that $\tilde{H}^g(x,\xi,t)$ and $\tilde{E}^g(x,\xi,t)$
are odd in $x$, whereas the other six distributions are even. Time
reversal invariance leads to the same constraints as in the case of
quarks, and Eqs.~(\ref{time}) to (\ref{special-hermit}) remain valid
with the subscripts $q$ changed into $g$. In particular, one also
finds that the first moment in $x$ of $\Td^g$ has to vanish, in
analogy to Eq.~(\ref{moment}).

To find a helicity representation for the gluon distributions we use
again the framework of light-cone quantization. We recall that in the
gauge $A^+=0$ we are working in one has $F^{+i} = \partial^+ A^i$. The
transverse components of the gluon potential $A^i$ are the ``good''
components of the field, and combinations with definite helicity are
projected out by contracting them with the two-dimensional
polarization vectors
\begin{equation}
\epsilon(+) = - \frac{1}{\sqrt{2}} 
              \left( \begin{array}{c} 1 \\ i \end{array} \right) ,
\hspace{3em}
\epsilon(-) = \frac{1}{\sqrt{2}}
              \left( \begin{array}{c} 1 \\ -i \end{array} \right) .
\end{equation}
In the region $\xi<x<1$, where the operator $F^{+i}(-\frac{1}{2}z)$ is
associated with an incoming and $F^{+j}(\frac{1}{2}z)$ with an
outgoing gluon, we then define helicity combinations
\begin{eqnarray}
  \label{helicity-elements-gluon}
A_{\lambda'\mu', \lambda\mu} &=&
\frac{1}{P^+} \int \frac{d z^-}{2\pi}\, e^{ix P^+ z^-}
\nonumber \\
&\times& \langle p',s'|\, 
     \epsilon^{i}(\mu')F^{+i}(-{\textstyle\frac{1}{2}}z)\, 
     F^{+j}({\textstyle\frac{1}{2}}z)\epsilon^{*j}(\mu) \,|p,s 
   \rangle\Big|_{z^+=0,\,\mathbf{z}_T=0} \, , 
\hspace{2em}
\end{eqnarray}
where summation over $i,j = 1,2$ is understood. One easily sees that
the matrix elements $A_{\lambda'\mu', \lambda\mu}$ which conserve
gluon helicity read exactly as those in Eq.~(\ref{no-flip-amplitudes})
with the superscript $q$ changed to $g$ in the distributions $H$,
$\tilde{H}$, $E$, $\tilde{E}$. For gluon helicity flip we find
\begin{eqnarray}
  \label{flip-amplitudes-gluon}
A_{++,+-} &=& \sqrt{1-\xi^2}\; \frac{t_0-t}{4 m^2} \left( \Tb^g
              + (1-\xi)\, \frac{\Tc^g + \Td^g}{2} \right) , 
\nonumber \\
A_{-+,--} &=& \sqrt{1-\xi^2}\; \frac{t_0-t}{4 m^2} \left( \Tb^g
              + (1+\xi)\, \frac{\Tc^g - \Td^g}{2} \right) , 
\nonumber \\
A_{++,--} &=& \epsilon\, (1-\xi^2)\, \frac{\sqrt{t_0-t}}{2m} 
\nonumber \\
          & & \times \left(\Ta^g + \
              \frac{t_0-t}{4 m^2}\, \Tb^g -
              \frac{\xi^2}{1-\xi^2}\, \Tc^g +
              \frac{\xi}{1-\xi^2}\, \Td^g \right) , 
\nonumber \\
A_{-+,+-} &=& - \epsilon\, (1-\xi^2)\, 
              \frac{\sqrt{t_0-t}^{\,3}}{8 m^3}\, \Tb^g \, .
\end{eqnarray}
The remaining helicity combinations are given by the parity relation
(\ref{parity}). Notice that all gluon helicity flip matrix elements go
to zero with a factor (\ref{prefactor}) in the collinear limit,
$t=t_0$, where angular momentum conservation requires
$\lambda-\mu=\lambda'-\mu'$. In other words, the gluon helicity flip
distributions for a spin $\frac{1}{2}$ target decouple from any
observable for collinear scattering.

\section{Compton scattering}
\label{sec:compton}

The generalized gluon helicity flip distributions appear in deeply
virtual Compton scattering, i.e., in the process $\gamma^* p \to
\gamma p$ at large photon virtuality $Q^2$, large c.m.\ energy, and
small squared momentum transfer $t$. This process is observable in
electroproduction, $e p \to e \gamma p$.

It is easy to see that to leading order in $1/Q$, the gluon helicity
flip distributions only contribute to the $\gamma^* p \to \gamma p$
amplitudes where the photon helicity is flipped by two units
\cite{Diehl:1997bu}. This is because, to that accuracy, the hard
scattering subprocess is collinear, so that the gluon helicity flip
needs to be compensated by the photons. Conversely, to leading order
in $1/Q$, it is only the helicity flip gluon distributions that appear
in the photon helicity flip amplitudes. Although only coming in at
order $\alpha_s$, these distributions thus provide the leading
contribution to photon helicity flip. This is different for the
helicity conserving gluon distributions (\ref{no-flip-gluon}). They
contribute at order $\alpha_s$ to the photon helicity conserving
amplitudes, which receive Born level contributions from the quark
distributions (\ref{no-flip-quark}). Given that the different photon
helicity amplitudes can be separated by the measurement of angular
distributions in the final state \cite{Diehl:1997bu}, one may thus be
able to experimentally access the gluon helicity flip distributions in
a rather direct and clean way.

\begin{figure}
   \begin{center}
        \leavevmode
        \epsfxsize=0.45\hsize
        \epsfbox{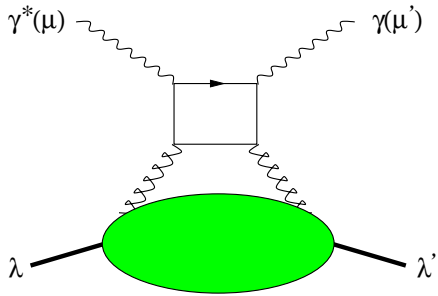}
   \end{center}
\caption{\label{fig:compton} A Feynman diagram for the photon helicity
flip amplitude in deeply virtual Compton scattering. The remaining
diagrams are obtained by appropriate permutations of the photon and
gluon lines. Greek letters label helicities, and the relevant
combinations are $(\mu',\mu) = (1,-1)$ and $(\mu',\mu) = (-1,1)$.}
\end{figure}

The Feynman diagrams involving gluon helicity flip (see
Fig.~\ref{fig:compton}) have been calculated by two groups
\cite{Hoodbhoy:1998vm,Belitsky:2000jk}. They give the relevant
$\gamma^* p \to \gamma p$ amplitudes to leading order in $1/Q$ as
\begin{eqnarray}
  \label{compton}
e^2 M_{\lambda'\mu',\lambda\mu} &=& 
  - \frac{\alpha_s}{2\pi} \sum_q e^2 e_q^2\, \int_{-1}^{1} dx\, 
  \frac{A_{\lambda'\mu',\lambda\mu}(x,\xi,t)}{(\xi-x-i\epsilon)
  (\xi+x-i\epsilon)}
\end{eqnarray}
for $(\mu',\mu) = (1,-1)$ and $(\mu',\mu) = (-1,1)$, where $\mu'$ and
$\mu$ in $M_{\lambda'\mu',\lambda\mu}$ denote the respective
helicities of the outgoing and incoming photon. The matrix elements
$A_{\lambda'\mu',\lambda\mu}$ are given by
Eqs.~(\ref{flip-amplitudes-gluon}) and (\ref{parity}). Our
polarization vectors for both photon states read $\epsilon(\pm) =
(0,\mp 1, i,0) /\sqrt{2}$, up to corrections in $1/Q$, where it is
understood that in our reference frame the photons are left-moving and
the protons right-moving.

Whether the gluon helicity flip distributions can be accessed
experimentally at a given value of $Q^2$, depends on their size. They
have to be large enough to compensate for the factor $\alpha_s /\pi$
in the leading-twist contribution~(\ref{compton}), which competes with
power suppressed terms that already start at zeroth order
in~$\alpha_s$. Unfortunately, nothing is presently known about the
size of gluon helicity flip distributions in the proton.

In which combinations the distributions $\Ta^g$, $\Tb^g$, $\Tc^g$,
$\Td^g$ appear in the cross section of $ep \to e\gamma p$, shall not
be studied here in detail. As an example we mention only their
contribution to the interference term between the Compton and the
Bethe-Heitler processes, where they give rise to a $\cos3\varphi$
angular distribution (for more details we refer to
\cite{Diehl:1997bu}). With unpolarized electrons and protons, they
appear in the combination\footnote{The accompanying global factor in
the cross section can be found in \protect\cite{Belitsky:2000jk},
where only the distributions $\Ta^g$ and $\Tc^g$ were considered.}
\begin{equation}
\frac{\sqrt{t_0-t}^{\,3}}{8 m^3} \left[ 
   \Ta^g\, F_2^{\phantom{g}} - \Tc^g\, F_1^{\phantom{g}} - 
   2 \Tb^g \left( F_1^{\phantom{g}} + \frac{t}{4m^2}\,
   F_2^{\phantom{g}}
\right) \right] \cos3\varphi ,
\end{equation}
where $F_1$ and $F_2$ are the Dirac and Pauli form factor of the
proton, respectively. It is amusing to note that the distribution
$\Td^g$ does not appear in this term, just as the parton helicity
conserving distributions $\tilde{E}^q$ and $\tilde{E}^g$ are absent in
the corresponding $\cos\varphi$ term, see Eq.~(30) of
\cite{Belitsky:2001gz}. $\Td^g$ is however present in the contribution
to the cross section from the Compton process alone, and also in the
interference term with the Bethe-Heitler process if the initial proton
is polarized.

\section{Summary}
\label{sec:sum}

We have given a complete set of generalized parton helicity flip
distributions for a spin $\frac{1}{2}$ target. Both in the quark and
gluon sector there are four independent distributions, i.e., two more
than previously considered in the literature
\cite{Hoodbhoy:1998vm,Belitsky:2000jk}.

The constraints of parity invariance are very similar for ordinary and
generalized parton distributions and for the helicity amplitudes of
two-particle elastic scattering, as is embodied in
Eq.~(\ref{parity}). This is not the case for invariance under time
reversal, which acts to reduce the number of independent helicity
amplitudes and of ordinary parton distributions, but \emph{not} of
generalized ones. This can be traced back to the dependence of
generalized parton distributions on the momentum fraction $\xi$ or,
more broadly speaking, to the fact that generalized distributions not
only depend on two independent hadron momenta, but also on a
light-cone direction. Time reversal symmetry determines the behavior
of generalized distribution under $\xi\to -\xi$ and fixes their
phase. For counting the independent generalized distributions of
targets with arbitrary spin, one can still use the analogy with
helicity amplitudes, but must at that stage only use the constraints
from parity invariance, not those from time reversal.

As a by-product of our investigation we found that time reversal
invariance reduces the number of form factors of the local quark
tensor current $\bar{\psi}\, \sigma^{\alpha\beta}\, \psi$ from four to
three. This illustrates that this symmetry also acts in slightly
different, although related ways on generalized parton distributions
and on elastic form factors.

At present we do not know a physical process that would give access to
generalized quark helicity distributions. Their gluonic counterparts
can be rather cleanly investigated in deeply virtual Compton
scattering, under the condition that they are large enough. This would
be extremely interesting because, as that they do not mix with the
quark sector under evolution, gluon helicity flip distributions should
provide a rather unique glimpse into the dynamics of glue in the
nucleon.

\section*{Acknowledgments} 

I gratefully acknowledge inspiring discussions with J.~Soffer and
O.~Teryaev, and useful correspondence with D.~M\"uller. I would like
to thank the Centre de Physique Th\'eorique at Marseille for a kind
invitation which initiated this work.

\appendix

\section*{Appendix}

In this appendix we study the form factor decomposition of the tensor
current (\ref{local-matrix}) for the proton, which is also needed for
the parameterization of quark helicity flip distributions. We will
classify the Dirac bilinears
\begin{equation}
  \label{decomposition}
  \bar{u}(p',\lambda')\, \Gamma_n^{\alpha\beta}\, u(p,\lambda)
\end{equation}
where $\Gamma_n^{\alpha\beta}$ is a matrix in Dirac space and
antisymmetric in the Lorentz indices $\alpha$ and $\beta$. By parity
invariance $\Gamma_n^{\alpha\beta}$ must be a Lorentz tensor, not a
pseudotensor. We do not use any restrictions from time reversal at
this point. Let us show that an independent set of bilinears can be
chosen as the one on the right-hand side of
Eq.~(\ref{flip-quark}). For this, we use the equations of motion for
spinors, i.e., the Gordon identities
\begin{eqnarray}
  \label{Gordon-1}
2m\, \bar{u}(p')\, \gamma^\alpha\, u(p) &=&
  \bar{u}(p')\, \left[ (p'+p)^\alpha + 
  i \sigma^{\alpha\beta} (p'-p)_\beta \right] u(p) , \\
  \label{Gordon-2}
2m\, \bar{u}(p')\, \gamma^\alpha\gamma_5\, u(p) &=&
  \bar{u}(p')\, \left[ \gamma_5 (p'-p)^\alpha + 
  i \sigma^{\alpha\beta}\gamma_5 (p'+p)_\beta \right] u(p) ,
\hspace{2em}
\end{eqnarray}
and furthermore
\begin{eqnarray}
  \label{motion-1}
0 &=& \bar{u}(p')\, \left[ (p'-p)^\alpha + 
      i \sigma^{\alpha\beta} (p'+p)_\beta \right] u(p) , \\
  \label{motion-2}
0 &=& \bar{u}(p')\, \left[ \gamma_5 (p'+p)^\alpha + 
      i \sigma^{\alpha\beta}\gamma_5 (p'-p)_\beta \right] u(p) .
\end{eqnarray}
Now we show that all possible parity even structures in
Eq.~(\ref{decomposition}) can be reduced to
\begin{eqnarray}
\Gamma_1^{\alpha\beta} = i \sigma^{\alpha\beta} , \hspace{4.1em}
&\hspace{2em}&
    \Gamma_2^{\alpha\beta} = 
                     P^\alpha \Delta^\beta - \Delta^\alpha P^\beta ,
\nonumber \\
\Gamma_3^{\alpha\beta} =  
                     \gamma^\alpha \Delta^\beta - 
                     \Delta^\alpha \gamma^\beta ,  &\hspace{2em}&
\Gamma_4^{\alpha\beta} = \gamma^\alpha P^\beta - 
                     P^\alpha \gamma^\beta .
\end{eqnarray}
The $\Gamma_n^{\alpha\beta}$ must be constructed out of invariants and
the vectors $P$ and $\Delta$. We treat all possible Dirac currents in
turn:
\begin{enumerate}
\item \textit{axial vector current
($\bar{u}\,\gamma^\delta\gamma_5\,u$):} We can use the Gordon identity
(\ref{Gordon-2}) and replace it by the pseudoscalar current and by
$\bar{u}\,i \sigma^{\gamma\delta}\gamma_5\,u$. Then go to 2 and 3.
\item \textit{pseudoscalar current ($\bar{u}\,\gamma_5\,u$):} Due to
parity invariance, this must be multiplied with a pseudotensor of rank
2. The only possible choice is $\epsilon^{\alpha\beta\gamma\delta}\,
\Delta_\gamma P_\delta$. Using Eq.~(\ref{motion-2}), we can eliminate
$\bar{u}\,\gamma_5 P_\delta\,u$ in favor of $\bar{u}\,i
\sigma^{\gamma\delta}\gamma_5\,u$. Then go to 3.
\item \textit{pseudotensor current ($\bar{u}\,i
\sigma^{\gamma\delta}\gamma_5\,u$):} can be replaced by the tensor
current using Eq.~(\ref{sigma-five}). Then go to 4.
\item \textit{tensor current ($\bar{u}\,i \sigma^{\gamma\delta}\,u$):}
If any of the indices $\gamma$ or $\delta$ is contracted with $P$ or
$\Delta$, we can express this current in terms of the scalar and
vector current using the identities (\ref{Gordon-1}) and
(\ref{motion-1}). The contraction of $\gamma$ and $\delta$ with the
$\epsilon$-tensor is not allowed by parity invariance, so that the
only possibility left gives $\Gamma_1^{\alpha\beta} = i
\sigma^{\alpha\beta}$.
\item \textit{vector current ($\bar{u}\,\gamma^\delta\,u$):} If the
index $\delta$ is contracted with $P$ or $\Delta$ we can use the
equations of motion to obtain the scalar current or zero. Again we
cannot contract with the $\epsilon$-tensor because of parity
invariance. The only possibilities left to form an antisymmetric
tensor of rank 2 are thus those in $\Gamma_3$ and $\Gamma_4$.
\item \textit{scalar current ($\bar{u}u$):} must be multiplied with an
antisymmetric tensor of rank 2. We cannot use the $\epsilon$-tensor
because of parity constraints, so the only possibility left leads to
$\Gamma_2$.
\end{enumerate}

As we discussed in Section~\ref{sec:quark}, time reversal invariance
forbids $\Gamma_4$ in the decomposition of the local current matrix
element $T^{\alpha\beta}$ in (\ref{local-matrix}), but not in the case
of the bilocal current (\ref{flip-quark}) defining the quark helicity
flip distributions. The second moment $\int dx\, x \Td^q(x,\xi,t)$ is
connected with the local matrix element
\begin{equation}
  \langle p',\lambda'|\, 
     \bar{\psi}(0)\, 
      \raisebox{0.2ex}{$\stackrel{\leftrightarrow}{\partial}{}^{\!\!+}$}
     i \sigma^{+i}\, \psi(0)\, 
  \,|p,\lambda \rangle .
\end{equation}
Now we do have four linearly independent tensors respecting the
constraints of both parity and time reversal invariance, namely
$P_{\phantom{1}}^+ \Gamma_1^{+i}$, $P_{\phantom{1}}^+ \Gamma_2^{+i}$ ,
$P_{\phantom{1}}^+ \Gamma_3^{+i}$, and $\Delta_{\phantom{1}}^+
\Gamma_4^{i+}$.

\end{document}